\documentclass[useAMS,usenatbib,letter]{mn2e}
\usepackage[dvips]{graphicx}

\usepackage{times}

\newcommand{\cm}{cm$^{-1}$}

\title[Experimental Ti I oscillator strengths]{Experimental Ti I oscillator strengths and their application to cool star analysis}
\author[R.J. Blackwell--Whitehead et al.]{R.J. Blackwell--Whitehead$^{1}$\thanks{E-mail:
r.blackwell@imperial.ac.uk (RBW)}, H. Lundberg$^{2}$, G. Nave$^3$,
J.C. Pickering$^{1}$, H.R.A. Jones$^4$,
\newauthor Y. Lyubchik $^5$, Y.V. Pavlenko$^5$, S. Viti$^6$\\
$^{1}$Blackett Laboratory, Imperial College, London, SW7 2BW,
UK\\
$^{2}$Department of Physics, Lund Institute of Technology, P.O.Box 118, 22100 Lund, Sweden\\
$^{3}$Atomic Physics Division, National Institute of Standards and
Technology, 100 Bureau Dr., Gaithersburg, MD 20899, USA\\
$^4$ Centre for Astrophysics Research, University of Hertfordshire, UK\\
$^5$ Main Astronomical Observatory, Kiev, Ukraine\\
$^6$ Department of Physics and Astronomy, University College
London, UK}
\begin{document}

\date{Accepted for publication 2006 September 26}

\pagerange{\pageref{firstpage}--\pageref{lastpage}} \pubyear{2006}

\maketitle

\label{firstpage}

\begin{abstract}
We report experimental oscillator strengths for 88 Ti I
transitions covering the wavelength range 465 to $3\:892$ nm, 67
of which had no previous experimental values. Radiative lifetimes
for thirteen energy levels, including the low energy levels 3d$^2
$(~$^3$F) 4s4p (~$^3$P) z~$^{5}$D$^{\circ}_{j}$, have been
measured using time resolved laser induced fluorescence. Intensity
calibrated Ti I spectra have been measured using Fourier transform
spectroscopy to determine branching fractions for the decay
channels of these levels. The branching fractions are combined
with the radiative lifetimes to yield absolute transition
probabilities and oscillator strengths. Our measurements include
50 transitions in the previously unobserved infrared region
$\lambda$ $>$ 1.0 $\mu$m, a region of particular interest to the
analysis of cool stars and brown dwarfs.
\end{abstract}


\section{INTRODUCTION}
Infrared stellar spectroscopy is becoming increasingly important
with the advent of better infrared (IR) detectors on ground--based
and satellite--borne spectrographs. However, at present the
analysis of expensively acquired IR astrophysical spectra is
restricted to the study of molecular bands and a small number of
atomic transitions in the near IR. A significant restriction on
the analysis is the lack of accurate laboratory determined atomic
oscillator strengths in the IR. The need for accurate IR atomic
oscillator strengths is particularly acute for the study of cool
stars and brown dwarfs whose energy distribution peaks in the IR.
Accurate oscillator strengths for specific neutral atoms are thus
needed to determine fundamental properties, such as effective
temperature, metallicity and surface gravity
\citep{Lyubchik04,Jones05a}.

The current laboratory atomic database for Ti I is particularly
poor in the infrared. The longest wavelength spectral line with a
measured oscillator strength is 1.06~$\mu$m \citep{Whaling77}. In
the wavelength region above 1.06~$\mu$m, the only available
oscillator strengths are derived from the semi-empirical
calculations of \cite{Kurucz95}. However, theoretical and
semi--empirical predictions of oscillator strengths are difficult
to calculate to the accuracy required for abundance determinations
(uncertainty of 10 to 20 per cent) especially for weak transitions
that may be the only useable lines for the analysis of
astrophysical spectra.

The most recent compilation of Ti I oscillator strengths is the
work by \cite{Martin88}, which includes the visible to near IR
study by \cite{Smith78} and
\cite{Blackwell82a,Blackwell82b,Blackwell83}. The Blackwell
oscillator strengths were re-examined by \cite{Grevesse89} who
recommended that they be increased by 0.056~dex, or 14 per cent.
Since then, \cite{Nitz98} have published experimentally measured
oscillator strengths with uncertainties of about 10 per cent for
some even parity levels. Experimental lifetime measurements on Ti
I have been carried out by \cite{Salih90,Lawler91,Lowe91} and
\cite{Rudolph82} using time resolved Laser Induced Fluorescence
(LIF). Despite the extensive work by \cite{Lawler91}, no
experimental lifetimes have been published for the low lying
levels, z~$^{5}$D$^{\circ}_{j}$. However, transitions from the
z$~^5$D$^{\circ}_{j}$ levels are some of the strongest features in
the IR Ti I spectrum, and our new measurements have been carried
out to obtain accurate laboratory oscillator strengths for these
transitions.

The transitions measured in the current work are a selection of
strong Ti I lines observed in the near IR that are of importance
to the analysis of ultracool dwarf stars, as outlined in
\cite{Lyubchik04}. We have measured branching fractions for 88
lines from 465.777 to $3\:867.179$ nm using high resolution
Fourier transform spectroscopy. To place the relative line
intensities on an absolute scale, we have measured 13 level
lifetimes using time resolved LIF and combined these with the
branching fractions to yield 88 oscillator strengths, 67 of these
being measured for the first time.

\section{Experimental measurements}
The oscillator strengths have been determined by combining
accurate branching fractions with level lifetime measurements. The
branching fraction, BF$_{ul}$, for a transition between upper
energy level $u$ and lower energy level $l$ is defined as the
ratio of its transition probability, A$_{ul}$, to the sum of the
transition probabilities of all the possible lines from the same
upper level:
\begin{equation}
BF_{ul}=\frac{A_{ul}}{\sum_{l}A_{ul}}=\frac{I_{ul}}{\sum_{l}I_{ul}}
\label{br_eq}\end{equation}

where I$_{ul}$ is the photon intensity of the spectral line. The
lifetime, $\tau_{u}$, of the upper level is the inverse sum of the
transition probabilities for all lines from the upper level:
\begin{equation}
 \tau_{u}=\frac{1}{\sum_{l}A_{ul}}
 \label{tau}
\end{equation}

The transition probability for a line is then defined in terms of
BF$_{ul}$ and $\tau_{u}$ as:
\begin{equation}
A_{ul}=\frac{BF_{ul}}{\tau_{u}}
\end{equation}

We have used the branching fractions and level lifetime technique
as outlined above to determine oscillator strengths; the
experimental details are described in the following sections.

\subsection{Branching fraction measurements}
The Ti I spectrum was measured using two spectrometers: the
visible to IR ($1\:800$ to $25\:000$ \cm ) region was recorded at
the National Institute of Standards and Technology (NIST) using
the NIST 2 m Fourier Transform Spectrometer (FTS) \citep{Nave97};
and in the visible to UV ($15\:000$ to $30\:000$ \cm ) at Imperial
College (IC) using the IC UV FTS \citep{Pickering02}. The light
source used for both the NIST and IC measurements was a water
cooled hollow cathode lamp (HCL) \citep{Danzmann88,Learner88}. A
pure Ti (99.99 per cent) cathode was used in the HCL, with either
Ar or Ne as a buffer gas, to produce a stable source for the Ti I
emission spectrum. The optimum running conditions for the Ti HCL
were found to be 100~Pa of Ar at a current of 500~mA, and 370~Pa
of Ne at a current of 500~mA. To check for self absorption in the
main high current measurements, additional low current
measurements were recorded with a Ne buffer gas at 200~mA and at a
pressure of 345~Pa. Branching fractions were determined using both
the spectra recorded with low and high HCL current as discussed by
\cite{Blackwell_05mn1gf}. The branching fractions agreed to within
the experimental uncertainties, indicating that no self-absorption
was present. A summary of the spectra recorded is given in Table
\ref{summary}.

\begin{table*}
 \centering
 \begin{minipage}{140mm}
  \caption{A summary of the Ti I spectra measured at IC and at
NIST.\label{summary}}
  \begin{tabular}{@{}rrlccccr@{}}
  \hline
   \multicolumn{3}{c}{Spectral range\footnote{The spectral range is the region over which the spectrum was intensity calibrated.}} & Resolution & Spectrometer & Buffer Gas & Current & Notes\\
   \multicolumn{3}{c}{(\cm)} & (\cm) & & & (mA) & \\
   \hline
  $1\:800$ & -- & $18\:500$ & 0.01 & NIST 2m FTS &  Ne & 500, 200 &\\
  $1\:800$ & -- & $18\:500$ & 0.01 & NIST 2m FTS &  Ar & 500 &\\
  $7\:500$ & -- & $25\:000$ & 0.01 & NIST 2m FTS &  Ar & 500 &\\
 $10\:000$ & -- & $22\:500$ & 0.01 & NIST 2m FTS &  Ne & 500 & KG3 filter\footnote{The KG3 filter is a near--IR short--pass filter.}\\
 $15\:000$ & -- & $30\:000$ & 0.03 & IC UV FTS &  Ne & 500 &\\
  \hline
\end{tabular}
\end{minipage}
\end{table*}

The intensity of the Ti I spectra was calibrated using two
tungsten intensity standard lamps; The IC tungsten lamp,
calibrated at the National Physical Laboratory (NPL) in the U.K.
over the spectral range $12\:500$ to $33\:000$ \cm\ (800 to 300
nm), has a minimum radiance uncertainty of $\pm$ 2.5 per cent (2
standard deviations). The NIST tungsten lamp was calibrated by
Optronics Laboratories\footnote{Certain trade names and products
are mentioned in the text in order to adequately identify the
apparatus used to obtain the measurements. In no case does such
identification imply recommendation or endorsement by NIST, IC or
the Lund Laser Centre. \label{disclaimer2}}in the U.S.A., and has
a radiance calibration uncertainty of $\pm$ 3.0 per cent over the
range $1\:600$ to $33\:000$ \cm\ ($6\:200$ to 300 nm). Spectra of
the tungsten lamp were recorded before and after each measurement
of the titanium spectrum using the same instrument parameters.
Further details of the intensity calibration are discussed by
\cite{Pickering01a}.

\begin{figure}
 \centering
  \includegraphics[width=84mm]{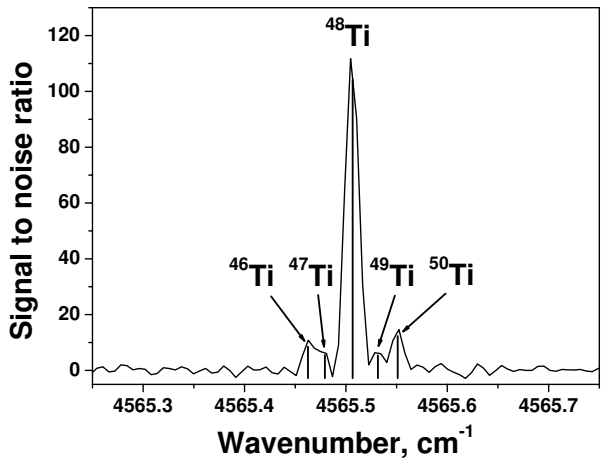}
  \caption{A section of the IR spectrum of Ti I. The a~$^{5}$P$_{2}$
  -- z~$^{5}$D$^{\circ}_{3}$ transition at 4565 \cm\ exhibits structure due to the
  five titanium isotopes $^{46}$Ti, $^{47}$Ti, $^{48}$Ti, $^{49}$Ti and $^{50}$Ti.}
 \label{isotopes}
\end{figure}

The most abundant isotope of titanium, $^{48}$Ti (73.7 per cent,
\cite{Rosman98}), was the major contributor to the Ti I line
profiles. In Fig. \ref{isotopes}, the individual isotope component
lines can be observed. The wavelength, intensity and
signal-to-noise ratio of each Ti I line was found by integrating
the intensity over the line profile and determining the centre of
gravity using computer code XGREMLIN \citep{Nave97}. The
calibrated integrated intensity was used to calculate the
branching fractions as described by equation \ref{br_eq}.

\subsection{Lifetime measurements}
The radiative lifetime measurements were carried out at the Lund
Laser Centre (LLC) using time resolved LIF. The experiment has
been described in detail by \cite{Xu03}, and only a brief
description is given here. The experimental apparatus consists of
two laser systems, the first of which, a Nd:YAG, laser is used to
produce free titanium atoms by laser ablation. The laser radiation
has a pulse energy of 2 to 10 mJ and is focused onto a rotating
titanium target located in a vacuum chamber that has a residual
pressure of about 10$^{-4}$ Pa. The second laser system consists
of a seeder injected Nd:YAG laser, a dye laser and a pulse
compressor. The Nd:YAG laser in the second system produces pulses
of approximately 8 ns in duration, which, after pulse compression,
provides a 1 ns pulse. The output pulse is then used to pump a dye
laser operating with a DCM dye. The output pulse from the dye
laser is tunable in the region 610 to 660 nm and can be
frequency--doubled or frequency--tripled as required. The tunable
range of the pulse can be extended by focusing the second harmonic
or third harmonic of the laser into a cell with hydrogen gas at
10$^{6}$ Pa to produce radiation of different stimulated Stokes
and anti-Stokes Raman components.

The free titanium atoms produced by laser ablation move across an
interaction area where the pulse from the second laser intersects
the titanium atoms and induces fluorescence. The two Nd:YAG lasers
are externally triggered so that the time between the ablation
pulse and the interaction of the excitation pulse with the free
atoms can be varied. The LIF is passed through a monochromator in
order to allow only one transition from the excited level to be
detected. The fluorescence from the z~$^5$D$^{\circ}_{j}$ levels
was detected with an IR microchannel plate photomultiplier tube
(PMT). All other levels have decay channels in the visible region
and hence a visible-region PMT was used. The photomultiplier
signal is recorded on an oscilloscope with a real time sampling
rate of 2 $\times$ 10$^{9}$ samples s$^{-1}$ and then processed by
a computer to determine the decay curve of the LIF signal.

\section{Results}
\subsection{Radiative lifetimes}\label{rad_sec}
The radiative lifetime, $\tau$, was determined for each level by
fitting exponential decay curves to approximately 20 measurements
and taking the average. Of the 13 level lifetimes measured, two
have previously published laboratory values \citep{Lawler91},
providing a comparison with our new measurements in Table
\ref{Radiative lifetimes}. The new laboratory lifetime values and
the previous values by \cite{Lawler91} agree to within the
experimental uncertainties of both measurements.

\begin{table*}
 \centering
 \begin{minipage}{140mm}
  \caption{Radiative lifetime measurements of Ti I levels. \label{Radiative lifetimes}}
  \begin{tabular}{@{}llrcrrrr@{}}
  \hline
   & & & & \multicolumn{3}{c}{Lifetime, $\tau$ (ns)}\\
   Configuration & Term & J & Energy (\cm)\footnote{Energy level values taken from \cite{Forsberg91}} & This work \footnote{Uncertainty in
parenthesis (nm) is one standard deviation.} & Previous\footnote{Previous laboratory measurements by \cite{Lawler91}} & Theory\footnote{Semi--empirical values calculated by \cite{Kurucz95}}\\
 \hline
                          &                     &   &               &           &         & \\
3d$^2$(~$^3$F)4s4p(~$^3$P)& z~$^{5}$D$^{\circ}$ & 1 & $18\:482.772$ &  500(50)  &         & 532\\
                          & z~$^{5}$D$^{\circ}$ & 2 & $18\:525.059$ &  530(50)  &         & 529\\
                          & z~$^{5}$D$^{\circ}$ & 3 & $18\:593.946$ &  550(50)  &         & 526\\
                          & z~$^{5}$D$^{\circ}$ & 4 & $18\:695.133$ &  520(50)  &         & 529\\
                          &                     &   &               &           &         & \\
3d$^2$(~$^3$F)4s4p(~$^3$P)& z~$^{3}$F$^{\circ}$ & 2 & $19\:322.984$ &           & 202(10) & 218\\
                          & z~$^{3}$F$^{\circ}$ & 3 & $19\:421.580$ &  205(10)  & 204(10) & 220\\
                          & z~$^{3}$F$^{\circ}$ & 4 & $19\:573.973$ &  210(10)  & 208(10) & 224\\
                          &                     &   &               &           &         & \\
3d$^2$(~$^3$F)4s4p(~$^3$P)& z~$^{3}$D$^{\circ}$ & 1 & $19\:957.852$ &           & 167(8)  & 160\\
                          & z~$^{3}$D$^{\circ}$ & 2 & $20\:006.042$ &           & 180(9)  & 173\\
                          & z~$^{3}$D$^{\circ}$ & 3 & $20\:126.060$ &           & 190(10) & 188\\
                          &                     &   &               &           &         & \\
3d$^2$(~$^3$F)4s4p(~$^3$P)& z~$^{3}$G$^{\circ}$ & 3 & $21\:469.487$ &           & 362(18) & 602\\
                          & z~$^{3}$G$^{\circ}$ & 4 & $21\:588.494$ &           & 336(17) & 524\\
                          & z~$^{3}$G$^{\circ}$ & 5 & $21\:739.707$ &           & 319(16) & 472\\
                          &                     &   &               &           &         & \\
3d$^2$4s(~$^4$F)5p        & r~$^{3}$F$^{\circ}$ & 2 & $43\:467.537$ & 14.0(1.0) &         & 22\\
                          & r~$^{3}$F$^{\circ}$ & 3 & $43\:583.354$ & 14.7(1.0) &         & 22\\
                          & r~$^{3}$F$^{\circ}$ & 4 & $43\:744.793$ & 13.7(1.0) &         & 23\\
                          &                     &   &               &           &         & \\
3d$^2$4s(~$^4$F)5p        &  $^{3}$G$^{\circ}$  & 3 & $44\:155.594$ & 15.9(1.0) &         & 68\\
                          &  $^{3}$G$^{\circ}$  & 4 & $44\:257.097$ & 15.8(1.0) &         & 68\\
                          &  $^{3}$G$^{\circ}$  & 5 & $44\:375.501$ &  4.2(0.3) &         & 26\\
                          &                     &   &               &           &         & \\
3d$^3$(~$^2$F)4p          & s~$^{3}$G$^{\circ}$ & 4 & $46\:837.983$ &   44(4)   &         & 8\\
 \hline
\end{tabular}
\end{minipage}
\end{table*}

For most of our new level lifetimes, the only previously available
values for comparison are the semi-empirical calculations of
\cite{Kurucz95}. For the z~$^5$D$^{\circ}_{j}$,
z~$^3$F$^{\circ}_{j}$ and z~$^3$D$^{\circ}_{j}$  levels, the
comparison between experimental lifetimes and the \cite{Kurucz95}
calculations agree to within a few per cent. However, for the
higher energy levels the experimental lifetimes are up to a factor
of 6 smaller than the theoretical lifetimes.

\subsection{Branching fractions and oscillator strengths}\label{bfosc}

Table \ref{osc} contains the branching fractions, BF, the
transition probabilities, A, and oscillator strengths, log(gf),
determined by our measurements. The new oscillator strengths are
compared with semi--empirical calculations of \cite{Kurucz95} and
previous experimental data where available. In Table \ref{osc}, it
can be seen that our experimental results and the calculated
values for the stronger transitions agree to within $\pm$50 per
cent for the majority of the transitions. The Ti I wavenumbers,
$\sigma$, in Table \ref{osc} have been calibrated using the
accurate Ar II line wavenumbers of \cite{Norlen73}. The wavenumber
calibration uncertainty in the Ti I wavenumbers is 0.005 \cm\ and
agrees with the values of \cite{Forsberg91} to within the
uncertainties of both measurements.

Since Ti I has a complex energy level structure, transitions from
a given upper level can extend from the mid--IR to the UV, but
some of these transitions will be too weak to be observed in our
experiment. To determine the contribution of weak, unobserved
transitions, it is possible to use theoretical predictions to
calculate a residual value (the sum of the transition
probabilities for all unobserved transitions from a given upper
level), e.g. the work on Fe II \citep{Nilsson00,Pickering01b} and
Ti II \citep{Pickering01a}. For the branching fractions
investigated in this paper, the residual values are about 0.1 to 3
per cent and are given at the end of each set of branching
fractions in column 5 of Table \ref{osc}.

The uncertainty of the oscillator strength measurements is
determined from the uncertainty in the level lifetimes and the
branching fractions. The uncertainty in integrated intensities is
determined from the intensity calibration of the observed spectra
and the uncertainty in cross--calibrating the line intensities
observed in different spectra. The uncertainty in the lifetime
measurements is one standard deviation. The uncertainty in the
oscillator strength is the sum in quadrature of the uncertainties
for both the branching fractions and lifetimes.

\begin{table*}
 \centering
 \begin{minipage}{240mm}
  \caption{Experimental and semi--empirical transition probabilities for Ti I \label{osc}}
  \begin{tabular}{@{}llrrrrrrrrccr@{}}
  \hline
  Upper & Lower & \multicolumn{2}{c}{Transition} & BF & \multicolumn{1}{c}{BF Unc.} & A\footnote{Transition probability} & \multicolumn{2}{c}{This work} & \multicolumn{3}{c}{Previous Experiment} & Expt -- Theory\\
  level & level & $\lambda_{vac}$ (nm) & $\sigma$ (\cm)\footnote{Wavenumber as observed in Fourier transform spectra.} &  & \multicolumn{1}{c}{(per cent)} & (10$^7$s$^{-1}$) &
  log(gf) & Unc.\footnote{The uncertainty in log(gf) expressed in dex, where $\pm$ 0.01 dex corresponds to approximately $\pm$ 2.5 per cent.} & log(gf) & Unc.$^{c}$ & Ref.\footnote{References: (1) \cite{Smith78}; (2)
\cite{Blackwell82a}; (3) \cite{Blackwell82b}; (4) \cite{Blackwell83}.} & (per cent)\footnote{Percentage difference between our new log(gf) values and the calculated values of \cite{Kurucz95}.}\\
 \hline
                        &                 &            &           &       &     &       &       &      &       &      &     &     \\
z~$^{5}$D$^{\circ}_{1}$ & a~$^{5}$F$_{1}$ &   838.5085 & 11925.938 & 0.306 &  5  &  6.11 & -1.71 & 0.04 &       &      &     &   8 \\
                        & a~$^{5}$F$_{2}$ &   841.4670 & 11884.008 & 0.641 &  5  & 12.82 & -1.39 & 0.04 &       &      &     &   8 \\
                        & a~$^{5}$P$_{1}$ &  2221.7299 &  4500.997 & 0.037 &  5  &  0.75 & -1.78 & 0.04 &       &      &     &  13 \\
                        & a~$^{5}$P$_{2}$ &  2245.0052 &  4454.333 & 0.010 &  6  &  0.19 & -2.36 & 0.05 &       &      &     & -13 \\
                        &                 &            &  Residual & 0.006 &     &       &       &      &       &      &     &     \\
                        &                 &            &           &       &     &       &       &      &       &      &     &     \\
z~$^{5}$D$^{\circ}_{2}$ & a~$^{5}$F$_{1}$ &   835.5458 & 11968.225 & 0.024 &  8  &  0.46 & -2.62 & 0.05 & -2.62 & 0.06 & (4) &  -5 \\
                        & a~$^{5}$F$_{2}$ &   838.4834 & 11926.294 & 0.258 &  5  &  4.86 & -1.59 & 0.04 &       &      &     &   0 \\
                        & a~$^{5}$F$_{3}$ &   842.8822 & 11864.054 & 0.635 &  5  & 11.97 & -1.20 & 0.04 & -1.19 & 0.05 & (3) &  -3 \\
                        & a~$^{5}$P$_{1}$ &  2201.0514 &  4543.283 & 0.018 &  5  &  0.34 & -1.91 & 0.04 &       &      &     &   5 \\
                        & a~$^{5}$P$_{2}$ &  2223.8927 &  4496.620 & 0.029 &  5  &  0.56 & -1.69 & 0.04 &       &      &     &   5 \\
                        & a~$^{5}$P$_{3}$ &  2262.7392 &  4419.422 & 0.003 & 12  &  0.05 & -2.74 & 0.06 &       &      &     & -20 \\
                        &                 &            &  Residual & 0.033 &     &       &       &      &       &      &     &     \\
                        &                 &            &           &       &     &       &       &      &       &      &     &     \\
z~$^{5}$D$^{\circ}_{3}$ & a~$^{5}$F$_{2}$ &   833.6681 & 11995.181 & 0.018 &  5  &  0.32 & -2.63 & 0.04 &       &      &     &  -5 \\
                        & a~$^{5}$F$_{3}$ &   838.0164 & 11932.941 & 0.191 &  5  &  3.48 & -1.59 & 0.04 &       &      &     &  -5 \\
                        & a~$^{5}$F$_{4}$ &   843.7971 & 11851.190 & 0.707 &  5  & 12.85 & -1.02 & 0.04 &       &      &     &  -5 \\
                        & a~$^{3}$P$_{2}$ &  1000.8408 &  9991.600 & 0.001 & 19  &  0.02 & -3.65 & 0.08 &       &      &     & -38 \\
                        & a~$^{5}$P$_{2}$ &  2190.3367 &  4565.508 & 0.037 &  5  &  0.68 & -1.47 & 0.04 &       &      &     &  10 \\
                        & a~$^{5}$P$_{3}$ &  2228.0101 &  4488.310 & 0.017 &  5  &  0.31 & -1.80 & 0.04 &       &      &     &   3 \\
                        &                 &            &  Residual & 0.029 &     &       &       &      &       &      &     &     \\
                        &                 &            &           &       &     &       &       &      &       &      &     &     \\
z~$^{5}$D$^{\circ}_{4}$ & a~$^{5}$F$_{3}$ &   830.9705 & 12034.122 & 0.006 &  7  &  0.12 & -2.97 & 0.05 &       &      &     & -15 \\
                        & a~$^{5}$F$_{4}$ &   836.6539 & 11952.374 & 0.108 &  5  &  2.07 & -1.71 & 0.04 &       &      &     &   3 \\
                        & a~$^{5}$F$_{5}$ &   843.7278 & 11852.163 & 0.807 &  5  & 15.52 & -0.83 & 0.04 &       &      &     &   3 \\
                        & a~$^{5}$P$_{2}$ &  2178.8892 &  4589.495 & 0.056 &  4  &  1.07 & -1.17 & 0.04 &       &      &     &  13 \\
                        &                 &            &  Residual & 0.024 &     &       &       &      &       &      &     &     \\
                        &                 &            &           &       &     &       &       &      &       &      &     &     \\
z~$^{3}$F$^{\circ}_{2}$ & a~$^{3}$F$_{2}$ &   517.5183 & 19322.987 & 0.855 & 10  & 42.34 & -1.07 & 0.04 & -1.06 & 0.05 & (2) &   5 \\
                        & a~$^{3}$F$_{3}$ &   522.1155 & 19152.852 & 0.055 & 13  &  2.71 & -2.26 & 0.06 & -2.23 & 0.05 & (2) &   3 \\
                        & b~$^{3}$F$_{2}$ &  1283.4950 &  7791.226 & 0.066 & 10  &  3.26 & -1.40 & 0.04 &       &      &     &  33 \\
                        & b~$^{3}$F$_{3}$ &  1301.5452 &  7683.176 & 0.011 & 10  &  0.52 & -2.18 & 0.04 &       &      &     &  40 \\
                        & a~$^{3}$G$_{3}$ &  2372.5504 &  4214.874 & 0.012 & 10  &  0.59 & -1.61 & 0.04 &       &      &     &  60 \\
                        &                 &            &  Residual & 0.002 &     &       &       &      &       &      &     &     \\
                        &                 &            &           &       &     &       &       &      &       &      &     &     \\
z~$^{3}$F$^{\circ}_{3}$ & a~$^{3}$F$_{2}$ &   514.8912 & 19421.577 & 0.094 & 10  &  4.58 & -1.89 & 0.05 & -1.95 & 0.05 & (2) &  18 \\
                        & a~$^{3}$F$_{3}$ &   519.4416 & 19251.443 & 0.787 & 10  & 38.60 & -0.96 & 0.04 & -0.95 & 0.05 & (2) &   5 \\
                        & a~$^{3}$F$_{4}$ &   525.3563 & 19034.700 & 0.016 & 21  &  0.81 & -2.63 & 0.08 & -2.39 & 0.05 & (2) & -53 \\
                        & a~$^{5}$F$_{4}$ &   788.7169 & 12678.821 & 0.003 & 14  &  0.14 & -3.05 & 0.06 &       &      &     &  45 \\
                        & b~$^{3}$F$_{2}$ &  1267.4562 &  7889.819 & 0.007 & 10  &  0.33 & -2.25 & 0.04 &       &      &     &  33 \\
                        & b~$^{3}$F$_{3}$ &  1285.0552 &  7781.767 & 0.067 & 10  &  3.29 & -1.25 & 0.04 &       &      &     &  43 \\
                        & b~$^{3}$F$_{4}$ &  1308.0847 &  7644.765 & 0.009 & 10  &  0.46 & -2.08 & 0.04 &       &      &     &  45 \\
                        & a~$^{3}$G$_{3}$ &  2318.3240 &  4313.461 & 0.001 & 14  &  0.04 & -2.70 & 0.06 &       &      &     &  43 \\
                        & a~$^{3}$G$_{4}$ &  2344.7878 &  4264.778 & 0.012 & 10  &  0.60 & -1.46 & 0.04 &       &      &     &  70 \\
                        &                 &            &  Residual & 0.004 &     &       &       &      &       &      &     &     \\
                        &                 &            &           &       &     &       &       &      &       &      &     &     \\
z~$^{3}$F$^{\circ}_{4}$ & a~$^{3}$F$_{3}$ &   515.3619 & 19403.839 & 0.073 &  8  &  3.51 & -1.90 & 0.04 & -1.96 & 0.05 & (2) &  15 \\
                        & a~$^{3}$F$_{4}$ &   521.1836 & 19187.098 & 0.808 &  6  & 38.85 & -0.85 & 0.04 & -0.82 & 0.05 & (2) &   3 \\
                        & a~$^{5}$F$_{5}$ &   785.4838 & 12731.008 & 0.006 &  8  &  0.28 & -2.64 & 0.04 &       &      &     &  50 \\
                        & b~$^{3}$F$_{3}$ &  1260.3724 &  7934.163 & 0.007 &  6  &  0.33 & -2.15 & 0.04 &       &      &     &  50 \\
                        & b~$^{3}$F$_{4}$ &  1282.5179 &  7797.162 & 0.088 &  6  &  4.22 & -1.03 & 0.04 &       &      &     &  55 \\
                        & a~$^{1}$G$_{4}$ &  1341.2775 &  7455.579 & 0.001 & 10  &  0.05 & -2.95 & 0.05 &       &      &     & -55 \\
                        & a~$^{3}$G$_{4}$ &  2263.8914 &  4417.173 & 0.001 & 10  &  0.04 & -2.56 & 0.04 &       &      &     &  75 \\
                        & a~$^{3}$G$_{5}$ &  2296.9602 &  4353.580 & 0.016 &  6  &  0.76 & -1.27 & 0.04 &       &      &     &  88 \\
                        &                 &            &  Residual & 0.001 &     &       &       &      &       &      &     &     \\
                        &                 &            &           &       &     &       &       &      &       &      &     &     \\
\hline
\end{tabular}
\end{minipage}
\end{table*}

\begin{table*}
  \contcaption{Experimental and semi--empirical transition probabilities for Ti I}
  \centering
 \begin{minipage}{240mm}
  \begin{tabular}{@{}llrrrrrrrrccr@{}}
  \hline
  Upper & Lower & \multicolumn{2}{c}{Transition} & BF & \multicolumn{1}{c}{BF Unc.} & A\footnote{Transition probability} & \multicolumn{2}{c}{This work} & \multicolumn{3}{c}{Previous Experiment} & Expt -- Theory\\
  level & level & $\lambda_{vac}$ (nm) & $\sigma$ (\cm)\footnote{Wavenumber as observed in Fourier transform spectra.} &  & \multicolumn{1}{c}{(per cent)} & (10$^7$s$^{-1}$) &
  log(gf) & Unc.\footnote{The uncertainty in log(gf) expressed in dex, where $\pm$ 0.01 dex corresponds to approximately $\pm$ 2.5 per cent.} & log(gf) & Unc.$^{c}$ & Ref.\footnote{References: (1) \cite{Smith78}; (2)
\cite{Blackwell82a}; (3) \cite{Blackwell82b}; (4) \cite{Blackwell83}.} & (per cent)\footnote{Percentage difference between our new log(gf) values and the calculated values of \cite{Kurucz95}.}\\
 \hline
                        &                 &            &           &       &     &       &       &      &       &      &     &     \\
z~$^{3}$D$^{\circ}_{1}$ & a~$^{3}$F$_{2}$ &   501.5585 & 19937.853 & 0.848 &  9  & 50.79 & -1.24 & 0.04 & -1.22 & 0.10 & (1) &  -8 \\
                        & a~$^{3}$P$_{0}$ &   869.4719 & 11501.235 & 0.050 &  9  &  2.99 & -1.99 & 0.04 & -2.24 & 0.06 & (4) &  18 \\
                        & a~$^{3}$P$_{1}$ &   873.7112 & 11445.429 & 0.038 &  9  &  2.30 & -2.10 & 0.04 & -2.33 & 0.06 & (4) &  20 \\
                        & b~$^{3}$F$_{2}$ &  1189.6135 &  8406.092 & 0.059 &  9  &  3.52 & -1.65 & 0.04 &       &      &     &  23 \\
                        & a~$^{3}$D$_{1}$ &  3893.5967 &  2568.319 & 0.001 & 15  &  0.07 & -2.33 & 0.06 &       &      &     &   2 \\
                        &                 &            &  Residual & 0.004 &     &       &       &      &       &      &     &     \\
                        &                 &            &           &       &     &       &       &      &       &      &     &     \\
z~$^{3}$D$^{\circ}_{2}$ & a~$^{3}$F$_{2}$ &   499.8490 & 20006.042 & 0.083 & 10  &  4.59 & -2.07 & 0.04 & -2.06 & 0.05 & (2) & -15 \\
                        & a~$^{3}$F$_{3}$ &   504.1363 & 19835.908 & 0.766 &  9  & 42.56 & -1.09 & 0.04 & -1.07 & 0.06 & (4) &  -5 \\
                        & a~$^{5}$F$_{2}$ &   745.8638 & 13407.273 & 0.002 & 16  &  0.08 & -3.46 & 0.07 &       &      &     &   8 \\
                        & a~$^{5}$F$_{3}$ &   749.3424 & 13345.034 & 0.001 & 20  &  0.07 & -3.56 & 0.08 &       &      &     &   5 \\
                        & a~$^{1}$D$_{2}$ &   784.2716 & 12750.686 & 0.002 & 16  &  0.11 & -3.28 & 0.07 &       &      &     &   3 \\
                        & a~$^{3}$P$_{1}$ &   868.5367 & 11513.619 & 0.065 &  9  &  3.59 & -1.69 & 0.04 & -1.88 & 0.06 & (4) &   8 \\
                        & a~$^{3}$P$_{2}$ &   876.9086 & 11403.697 & 0.022 &  9  &  1.20 & -2.16 & 0.04 &       &      &     &   7 \\
                        & b~$^{3}$F$_{2}$ &  1180.0409 &  8474.283 & 0.010 &  9  &  0.54 & -2.25 & 0.04 &       &      &     &  18 \\
                        & b~$^{3}$F$_{3}$ &  1195.2815 &  8366.230 & 0.047 &  9  &  2.63 & -1.55 & 0.04 &       &      &     &  10 \\
                        & a~$^{3}$D$_{2}$ &  3872.6956 &  2582.181 & 0.001 & 11  &  0.08 & -2.07 & 0.05 &       &      &     &  25 \\
                        &                 &            &  Residual & 0.002 &     &       &       &      &       &      &     &     \\
                        &                 &            &           &       &     &       &       &      &       &      &     &     \\
z~$^{3}$D$^{\circ}_{3}$ & a~$^{3}$F$_{3}$ &   501.1042 & 19955.929 & 0.043 & 13  &  2.24 & -2.23 & 0.06 & -2.20 & 0.05 & (2) & -25 \\
                        & a~$^{3}$F$_{4}$ &   506.6065 & 19739.185 & 0.788 &  9  & 41.49 & -0.95 & 0.04 & -0.94 & 0.05 & (2) &  -3 \\
                        & a~$^{5}$F$_{4}$ &   747.2000 & 13383.297 & 0.002 & 12  &  0.12 & -3.15 & 0.05 &       &      &     &  35 \\
                        & a~$^{1}$D$_{2}$ &   776.9583 & 12870.704 & 0.003 & 13  &  0.14 & -3.06 & 0.06 &       &      &     &  18 \\
                        & a~$^{3}$P$_{2}$ &   867.7757 & 11523.715 & 0.088 &  9  &  4.61 & -1.44 & 0.04 & -1.61 & 0.06 & (4) &   3 \\
                        & b~$^{3}$F$_{3}$ &  1178.3772 &  8486.247 & 0.009 &  9  &  0.45 & -2.18 & 0.04 &       &      &     &  13 \\
                        & b~$^{3}$F$_{4}$ &  1197.7131 &  8349.245 & 0.051 &  9  &  2.69 & -1.39 & 0.04 &       &      &     &  13 \\
                        & a~$^{3}$D$_{3}$ &  3867.1786 &  2585.865 & 0.001 & 10  &  0.08 & -1.93 & 0.04 &       &      &     &  10 \\
                        &                 &            &  Residual & 0.016 &     &       &       &      &       &      &     &     \\
                        &                 &            &           &       &     &       &       &      &       &      &     &     \\
z~$^{3}$G$^{\circ}_{3}$ & a~$^{3}$F$_{2}$ &   465.7772 & 21469.493 & 0.810 & 11  & 22.38 & -1.29 & 0.05 & -1.28 & 0.05 & (2) &  60 \\
                        & b~$^{3}$F$_{2}$ &  1006.2662 &  9937.728 & 0.028 & 11  &  0.78 & -2.08 & 0.05 &       &      &     &  30 \\
                        & b~$^{3}$F$_{3}$ &  1017.3276 &  9829.676 & 0.002 & 16  &  0.06 & -3.17 & 0.07 &       &      &     &   5 \\
                        & a~$^{1}$G$_{4}$ &  1069.3930 &  9351.100 & 0.002 & 18  &  0.05 & -3.22 & 0.08 &       &      &     & -70 \\
                        & a~$^{3}$G$_{3}$ &  1571.9872 &  6361.375 & 0.088 & 10  &  2.43 & -1.20 & 0.05 &       &      &     &  23 \\
                        & a~$^{3}$G$_{4}$ &  1584.1108 &  6312.690 & 0.011 & 11  &  0.29 & -2.12 & 0.05 &       &      &     &  28 \\
                        & a~$^{3}$D$_{2}$ &  2472.0023 &  4045.304 & 0.001 & 23  &  0.01 & -3.05 & 0.09 &       &      &     & 500 \\
                        & a~$^{3}$H$_{4}$ &  2913.5219 &  3432.272 & 0.028 & 10  &  0.76 & -1.17 & 0.05 &       &      &     &  28 \\
                        & b~$^{1}$G$_{4}$ &  3142.7448 &  3181.932 & 0.003 & 11  &  0.09 & -2.03 & 0.05 &       &      &     & 353 \\
                        &                 &            &  Residual & 0.028 &     &       &       &      &       &      &     &     \\
                        &                 &            &           &       &     &       &       &      &       &      &     &     \\
z~$^{3}$G$^{\circ}_{4}$ & a~$^{3}$F$_{3}$ &   466.8891 & 21418.362 & 0.818 & 10  & 24.33 & -1.15 & 0.05 & -1.13 & 0.05 & (2) &  50 \\
                        & b~$^{3}$F$_{3}$ &  1005.1583 &  9948.682 & 0.029 & 10  &  0.86 & -1.93 & 0.04 &       &      &     &  30 \\
                        & b~$^{3}$F$_{4}$ &  1019.1939 &  9811.676 & 0.002 & 15  &  0.06 & -3.10 & 0.06 &       &      &     &  15 \\
                        & a~$^{3}$G$_{3}$ &  1543.1200 &  6480.377 & 0.004 & 11  &  0.13 & -2.38 & 0.05 &       &      &     &  18 \\
                        & a~$^{3}$G$_{4}$ &  1554.8008 &  6431.692 & 0.086 & 10  &  2.55 & -1.08 & 0.04 &       &      &     &  30 \\
                        & a~$^{3}$G$_{5}$ &  1570.3276 &  6368.098 & 0.010 & 11  &  0.29 & -2.01 & 0.05 &       &      &     &  40 \\
                        & a~$^{3}$H$_{4}$ &  2815.8907 &  3551.274 & 0.001 & 13  &  0.03 & -2.53 & 0.06 &       &      &     &  -7 \\
                        & a~$^{3}$H$_{5}$ &  2900.8846 &  3447.224 & 0.029 & 10  &  0.86 & -1.01 & 0.04 &       &      &     &  45 \\
                        &                 &            &  Residual & 0.022 &     &       &       &      &       &      &     &     \\
                        &                 &            &           &       &     &       &       &      &       &      &     &     \\
z~$^{3}$G$^{\circ}_{5}$ & a~$^{3}$F$_{4}$ &   468.3220 & 21352.832 & 0.832 & 10  & 26.09 & -1.03 & 0.04 & -1.01 & 0.05 & (2) &  43 \\
                        & b~$^{3}$F$_{4}$ &  1003.7242 &  9962.896 & 0.033 & 10  &  1.03 & -1.77 & 0.04 &       &      &     &  38 \\
                        & a~$^{3}$G$_{4}$ &  1519.0857 &  6582.907 & 0.004 & 11  &  0.13 & -2.32 & 0.05 &       &      &     &  33 \\
                        & a~$^{3}$G$_{5}$ &  1533.9037 &  6519.314 & 0.091 & 10  &  2.84 & -0.96 & 0.04 &       &      &     &  28 \\
                        & a~$^{3}$H$_{5}$ &  2778.9797 &  3598.443 & 0.001 & 12  &  0.03 & -2.40 & 0.05 &       &      &     &  10 \\
                        & a~$^{3}$H$_{6}$ &  2819.1750 &  3547.137 & 0.034 & 10  &  1.07 & -0.85 & 0.04 &       &      &     &  58 \\
                        &                 &            &  Residual & 0.006 &     &       &       &      &       &      &     &     \\
                        &                 &            &           &       &     &       &       &      &       &      &     &     \\
\hline
\end{tabular}
\end{minipage}
\end{table*}

It can be seen that our results are consistent with previous
experimental measurements for both the weaker and stronger
transitions. Of the previous 21 laboratory determined oscillator
strengths in Table \ref{osc}, 16 agree to within the experimental
uncertainties of both measurements.  Although most of our
measurements agree well with the oscillator strengths of
\cite{Blackwell83} that have been revised by \cite{Grevesse89},
the four IR transitions from the z~$^{3}$D$^{\circ}_{j}$ levels
measured by \cite{Blackwell83} are 56 per cent (0.25 dex) weaker
than our measurements. This may be the result of non--thermal
equilibrium effects in the King furnace technique used by
\cite{Blackwell83} or an inaccuracy in the measured furnace
temperature for the near IR oscillator strengths. It was noted by
\cite{Blackwell83} that these 4 lines gave a solar titanium
abundance that was 0.15 dex (40 per cent) higher than the lines at
shorter wavelengths, and they attributed this to blends in the
solar spectrum. Although their values agreed well with
\cite{Kurucz75}, the more recent calculations of \cite{Kurucz95}
show much better agreement with our data than those of
\cite{Blackwell83}. Our oscillator strengths for the transitions
from the z~$^{3}$D$^{\circ}_{j}$ levels agree with the
calculations of \cite{Kurucz95} and to within uncertainties of the
experimental measurements of \cite{Smith78} and
\cite{Blackwell82a}. Therefore we believe that the near IR
oscillator strengths of \cite{Blackwell83} should be reduced by
0.25 dex.

\subsection{Comparison with cool star spectra}
\begin{figure*}
\begin{minipage}{168mm}
 \centering
  \includegraphics[width=168mm]{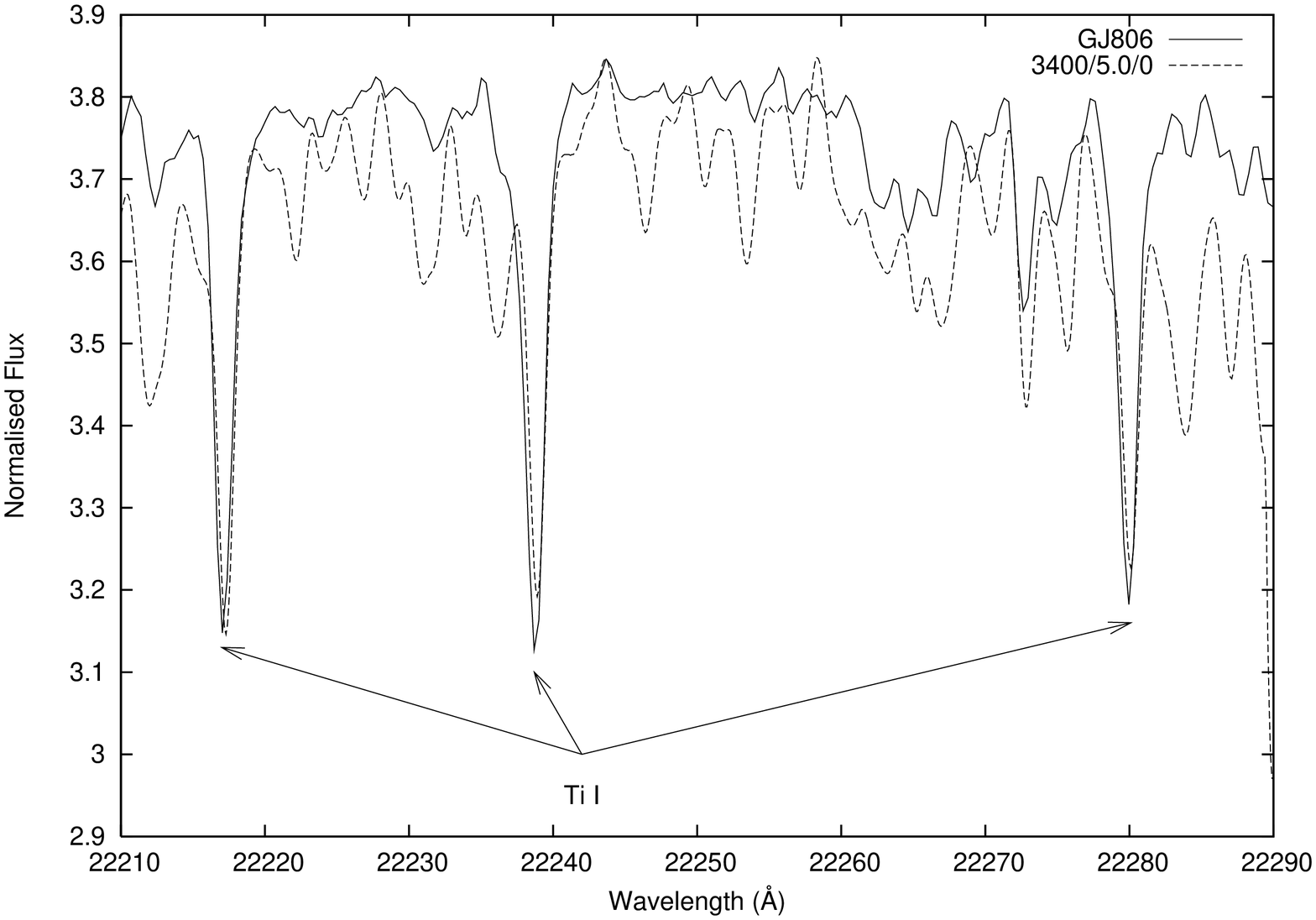}
  \caption{A section of an observed infrared spectrum from the cool dwarf star
GJ806 is compared with a $3\:400$ K, solar metallicity, log(g) $=$
5.0 theoretical spectrum (3400$/$5.0$/$0) using the new water line
list from Barber et al. (2006), and computed from the model
structure provided by Haushildt et al. (1999). The three labelled
Ti I lines are all ones whose measured oscillator strengths have
been described for the first time in this paper. They are all
denoted as high priority lines for the precision study of cool
stars by Lyubchik et al. (2004). }
 \label{starspec}
\end{minipage}
\end{figure*}

The new Ti I oscillator strengths presented in Table \ref{osc}
include transitions listed by \cite{Lyubchik04} as important for
the analysis of ultra cool dwarf stars. In total, five of the new
oscillator strengths are listed by \cite{Lyubchik04} as first
priority transitions, thirteen are listed as second priority and
three as third priority. In Fig. \ref{starspec}, we show an
observed spectrum of the M2 dwarf GJ806. The spectrum was acquired
by \cite{Doppmann06} using the \textit{NIRSPEC} instrument on the
Keck Telescope and has a resolution of $17\:000$. The theoretical
spectrum was generated by the WITA6 programme \citep{Pavlenko00},
and used the NEXTGEN model structures supplied by
\cite{Hauschildt99}. The theoretical spectrum was computed with a
wavelength step of 0.001 nm and convolved with a Gaussian to match
the instrumental broadening. The Ti I lines identified in Fig.
\ref{starspec} have also been measured in the laboratory to
determine accurate oscillator strengths, see Table \ref{osc}. The
experimental values are the first laboratory determined oscillator
strengths for these transitions, which are all identified as
priority 1 lines by \cite{Lyubchik04}. This prioritization is an
indication that they are amongst the relatively few identifiable
atomic lines that are predicted to be observable and suitable for
the high precision analysis of cool stars and brown dwarfs. Whilst
the Ti I lines are clearly the strongest features in the observed
spectrum and are reasonably well fitted in Fig. \ref{starspec}, it
can also be seen that the continuum is rather poorly described. A
better fit of the continuum would benefit from a line-by-line
identification of the species responsible. In particular, scrutiny
should be made of the large numbers of molecular water vapour
lines expected to be present in this region as indicated by
\cite{Jones05b}. A detailed analysis of this observational
spectrum, and others that we are acquiring in wavelength regions
containing Ti I lines with experimentally determined oscillator
strengths, will be the subject of a further paper (Lyubchik et
al., in preparation).

\section{Summary}
Experimental oscillator strengths have been determined for 88
lines of Ti I in the visible and IR regions of the spectrum. Sixty
six of these transitions have no previous experimental oscillator
strengths. The uncertainty in the oscillator strengths is 10 to 15
per cent for the stronger transitions, and for all oscillator
strengths in Table \ref{osc} the uncertainty has been reduced to
$<$ 25 per cent.

In the IR region $\lambda$ $>$ 1.0 $\mu$m, 50 new oscillator
strengths have been measured, including those identified as high
priority by \cite{Lyubchik04}. We show that some of these features
are readily identifiable in cool star spectra and are likely to be
of substantial interest for measurement of effective temperatures,
gravities and metallicities. It can be seen that oscillator
strengths for transitions in the infrared region $\lambda$ $>$ 1.0
$\mu$m show a marked deviation from the semi--empirical
calculations of \cite{Kurucz95}, which were the only available
data for these lines prior to our work. Whilst theoretical
calculations provide important data in regions of the spectrum
where experimental data are not available, it is important to note
that significant differences between theoretical calculations and
experimental measurements do occur. In particular, for weaker
transitions, caution should be exercised when relying solely on
theoretical calculations for oscillator strengths. This is of
particular importance for the analysis of cool stars and brown
dwarfs where visible transitions are too weak to be of diagnostic
value and oscillator strengths in the IR are required.

\section*{Acknowledgments}
RBW and JCP gratefully acknowledge the financial support of PPARC
of the UK and the Leverhulme Trust. GN acknowledges the financial
support of the National Aeronautical and Space Administration
under inter-agency agreement W-10,255. HL acknowledges the support
of the EU-TMR access to Large-Scale Facility Programme (contract
RII3-CT-2003-506350). We are also very grateful to Kevin Covey for
making the spectrum of GJ806 available to us.

\bsp

\label{lastpage}

\end{document}